\documentclass[twocolumn,brazil,english,pra,aps,amssymb,amsfonts,floatfix]{revtex4}
\usepackage[T1]{fontenc}
\usepackage[latin1]{inputenc}
\usepackage{float}
\usepackage{amsmath}
\usepackage{graphicx}
\usepackage{amssymb}

\makeatletter
\usepackage{amscd}
\usepackage{bm}
\usepackage{babel}
\makeatother

\begin{document}

\title{Coherence and Entanglement in a Stern-Gerlach experiment}

\author{T. R. Oliveira and A. O. Caldeira}

\affiliation{Departamento de F{\'\i}sica da Mat{\'e}ria Condensada, Instituto de F{\'\i}sica
Gleb Wataghin,Universidade Estadual de Campinas}

\affiliation{Caixa Postal 6165, Campinas, SP, CEP 13083-970, Brazil}

\begin{abstract}
We give a simple example of the tight connection between
entanglement and coherence for pure bipartite systems showing the
double role played by entanglement; it allows for the creation of
superpositions of macroscopic objects but at the same time makes
subsystems lose their quantum mechanical coherence. For this we
study the time evolution of the spin coherence in the
Stern-Gerlach (SG) experiment. We also show that, contrary to the
naive intuition, the spin coherence is lost before the two beams
become separated in the spatial coordinates.
\end{abstract}

\email{tro@ifi.unicamp.br}

\maketitle

\section{INTRODUCTION}

The SG experiment is viewed as the standard evidence of the
quantum nature of the spin of a particle. When a beam of spin 1/2
particles, in the eigenstate $|+\rangle$ of $S_{x}$, goes through
a variable magnetic field in the $\hat{z}$ direction and the
emerging particles are detected on a screen, one observes the
presence of two distinct peaks corresponding to the spins in the
positive and negative  $\hat{z}$ direction.

Besides furnishing the evidence of the spin quantization, this
experiment is considered the paradigm of the measurement process
being the simplest example of coherence and entanglement; two
features of quantum mechanics of which one finds no analogue in
the classical world.

In the SG experiment the particle enters the magnet in a pure
state, for example, the eigenstate $|+\rangle_{x}$ of $S_{x}$, and
for the effect of  measurement of the $S_{z}$ spin component it is
described as a coherent superposition of the eigenstates of
$S_{z}$. Later on, as an effect of the interaction with the non-
uniform part of the magnetic field, these spin degrees of freedom
entangle with the spatial coordinates generating the state
$|\psi\left(t\right)\rangle=\alpha\left|+\right\rangle
\left|\varphi_{+}\left(t\right)\right\rangle
+\beta\left|-\right\rangle
\left|\varphi_{-}\left(t\right)\right\rangle $. In the end, as a
result of the measuring process, the quantum state collapses into
one of these eigenstates.

Coherence can be observed by measuring the spin in the $x$
direction to obtain $\langle S_{x}\rangle$. We know that spin
coherence is lost as the state evolves in time and the two spatial
parts become orthogonal. Since the measurement process naturally
involves the partial trace over the spatial part of the initially
pure global state the remaining spin part becomes a mixture.

But how is this spin coherence lost in the SG experiment? Is it
lost just when the two beams are far away? These are interesting
questions that address the very origin of the entanglement between
the states representing different degrees of freedom of the
particles.

Here we intend to study  the evolution of the spin coherence in
the SG experiment and how the entanglement between the spin and
spatial coordinates is affected in the course of time to show the
double role played by entanglement and coherence.

\section{STERN-GERLACH MODEL}

We are going to consider the usual SG model, where we just take
into account one direction of the magnetic field. For discussions
on these approximations see \cite{Platt,key-2}. Within this model
our Hamiltonian is \[ H=\frac{p^{2}}{2m}-f\sigma_{z}z,\] \\ where
$f=\mu\left(\partial B/\partial z\right)$ and we are not
considering the uniform part of the magnetic field, that is just
responsible for the spin precession. This Hamiltonian takes us to
the following propagator\cite{Schulman}
\begin{eqnarray*}
K_{ss'}\left(z,t;z',0\right) & = & \left\langle S\right.
\left|S'\right\rangle \sqrt{\frac{m}{2\pi i\hbar t}}
exp\left\{ \frac{i}{\hbar} \left[ \frac{m}{2t}\left(z-z'\right)^{2}-\right.\right.\\
 &  & \left. \left.\frac{m}{t}\triangle z\left(t\right)
 \left(z-z'\right)-\triangle p\left(t\right)z'-
 \frac{f^{2}}{24 m}t^{3} \right] \right\}
\end{eqnarray*}
with
\[\triangle p(t)=ft,\,\,\,\,\triangle z(t)=\frac{ft^{2}}{2m},\]
and
\[\overline{\triangle z(t)}=\frac{t\triangle p(t)}{m}-\triangle z(t).\]
This propagator can be used to perform the temporal evolution of
the physical state in the Schr{\"o}dinger prescription and we
employ the superposition below as the initial state:

\selectlanguage{brazil}

\begin{equation}
\left|\psi\right\rangle =\left(\alpha\left|+\right\rangle
+\beta\left|-\right\rangle \right)
\otimes\left|\varphi\right\rangle .
\end{equation}
Here $\left|\varphi\right\rangle $ is the spatial part of the
physical state which is not initially entangled with the spin.
Considering this initial spatial part as a wave packet with
minimum uncertainty we have
\begin{equation} \left\langle
z\right.\left|\varphi\right\rangle
=\varphi\left(z,0\right)=\frac{1}{\sqrt{\sqrt{2\pi}\sigma}}\,\,
e^{-\frac{z^{2}}{4\sigma^{2}}},
\end{equation}
which is a Gaussian packet with width $\sigma$ and centered at the
origin. Performing the temporal evolution we got

\begin{equation}
|\psi\left(t\right)\rangle=\alpha\varphi_{+}\left(t\right)\left|+\right\rangle
+ \beta\varphi_{-}\left(t\right)\left|-\right\rangle,
\end{equation}
with
\begin{eqnarray}
\varphi_{\pm}\left(z,t\right) & = & \frac{e^{i\theta\left(t\right)}}{\sqrt{\sigma\left(t\right)\sqrt{2\pi}}}
exp\left\{ -\frac{1}{4\sigma\left(t\right)^{2}}\left[z\mp\overline{\triangle z}\left(t\right)\right]^{2}+\right.
\nonumber \\
 & + & i\left[\frac{m}{2\hbar t}z^{2}\pm\frac{m}{\hbar t}\triangle z\left(t\right)z+
 \frac{f^{2}t^{3}}{24m\hbar^{2}}+\right.\nonumber \\
 & - & \left.\left.\frac{m}{2\hbar t}\left(\frac{\sigma}{\sigma\left(t\right)}\right)^{2}
 \left(z\mp\overline{\triangle z}\left(t\right)\right)^{2}\right]\right\} \label{eq:4}
\end{eqnarray}
and
\[\sigma\left(t\right)^{2}=\sigma^{2}+\left(\frac{\hbar t}{2m\sigma}\right)^{2}.\]

Since the function $\theta\left(t\right)$ appears only as a time
dependent  exponent of a complex phase it will not contribute to
the evaluation of the probability amplitudes.

As expected, after having interacted with the magnetic field the
spatial and spin degrees of freedom of the particle are now
entangled. Within a time interval $t$ the spatial part of the wave
function is a Gaussian whose  center follows the classical
trajectory with a time dependent width given by
$\sigma\left(t\right)$.

\section{Coherence and entanglement}

As we are  interested in the spin coherence we should trace over
the spatial coordinates and look at the off-diagonal elements of
the density operator in the spin space which reads, for $\alpha=\beta=1/\sqrt{2}$,
\[\rho_{+-}\left(t\right)=\int dz\,\varphi_{+}\left(z,t\right)\varphi_{-}^{\star}\left(z,t\right).\]
\begin{eqnarray*}
\rho_{+-}\left(t\right) & = & exp\left\{ -\frac{1}{2}\left[\frac{1}{2}
\frac{\triangle p\left(t\right)}{\hbar/2\sigma}\left(\frac{\sigma}{\sigma\left(t\right)}+
\frac{\sigma\left(t\right)}{\sigma}\right)\right]^{2}+\right.\\
 & - & \left.\frac{1}{2}\left(\frac{\overline{\triangle z}\left(t\right)}
 {\sigma\left(t\right)}\right)^{2}\right\}.
\end{eqnarray*}
In this expression we can see that we have two terms contributing
to the loss of coherence. The first one is the distance between
the centers of the packets in momentum space, as measured in
unities of the spread of the packets in this space, namely
$\hbar/2\sigma$. The second term also measures the distance
between the packets but in coordinate space instead. In this case
this measure is taken using the spread of the packet in coordinate
space, $\sigma\left(t\right)$, as the standard. We are interested
in knowing how long it takes for the spin coherence to be lost and
what is the contribution of each of those two above-mentioned
terms . Defining a ``decoherence'' time, $\tau$, as the time scale
within  which the off-diagonal element decay to $1/e$, we can show
that\[
\tau=\sqrt{\frac{2\sqrt{2}m\sigma}{f}}\left[-\frac{2\sqrt{2}fm\sigma^{3}}{\hbar^{2}}+
\sqrt{1+\frac{8f^{2}m^{2}\sigma^{6}}{\hbar^{4}}}\right]^{1/2}.\]

As this expression does not tell us much if it is analyzed in its
full extent we are going to do it for two particular limits,
\[
\frac{8f^{2}m^{2}\sigma^{6}}{\hbar^{4}}\left\{
\begin{array}{c}
\gg1\,\,\,\textrm{first\, case}\\
\ll1\,\,\,\textrm{second\, case}
\end{array}\right.
\]\\
In the first case we  approximate the decoherence time by
$\tau_{1}=\hbar/\sqrt{2}f\sigma$ whereas in the second case we do
it by $\tau_{2}=\sqrt{2\sqrt{2}m\sigma/f}$.\\
Now we want to investigate how the separation between the packets
evolves during the loss of coherence. This distance, in the
coordinate representation,  is given by the fraction
$\overline{\triangle z}\left(t\right)/\sigma\left(t\right)$ which
assumes the following values
\[ \frac{\overline{\triangle z}\left(\tau_{1}\right)}{\sigma\left(\tau_{1}\right)}\approx
\frac{1}{\sqrt{2}}\frac{\hbar^{2}}{2\sqrt{2}fm\sigma^{3}}\ll1\]
and
\[\frac{\overline{\triangle z}\left(\tau_{2}\right)}{\sigma\left(\tau_{2}\right)}\approx
\sqrt{\frac{2\sqrt{2}fm\sigma^{3}}{\hbar^{2}}}\ll1.\]

We see that in both cases the packets are not well separated in
space when the coherence is lost, and, therefore, the separation
of the packets in momentum space must be responsible for the loss
of spin coherence in the SG experiment. This can be viewed in the
figures 1, 2 and 3, where we plot the loss of coherence, as given
by the off-diagonal elements of the density operator in the spin
space, and the probability amplitudes for the two spatial parts.
These plots were made using the typical values of a SG experiment
\cite{Phys Today}: $m=1,8\times10^{-25}Kg$ (cooper atom mass),
$\partial B/\partial z=10^{3}T/m$ and $\sigma=10^{-5}m$.

One should also notice that in the first case the spread of the
packets in coordinate space is not relevant, $\sigma(t) \approx
\sigma$, whereas in the second case it has to be taken in account
since $\sigma(t) \gg \sigma$. Another way to understanding why the
momentum separation is responsible for the loss of coherence is to
note that in the beginning, when $\sigma(t) \approx \sigma$,
\begin{equation}
\frac{\overline{\triangle z}(t)}{\sigma(t)}\approx \frac{f}{2m\sigma}t^2
\end{equation}
while at long times, when $\sigma(t) \gg \sigma$, one has
\begin{equation}
\frac{\overline{\triangle z}(t)}{\sigma(t)}\approx
\frac{f\sigma}{\hbar}t.
\end{equation}
Nevertheless, the momentum separation is given by
\begin{equation}
\frac{\triangle p(t)}{\hbar/2\sigma}=\frac{2f\sigma}{\hbar}t
\end{equation}
being always linear in $t$. The only possibility the space
separation between the packets exceeds their momentum separation
is for long times when the space separation is no longer quadratic
but linear instead.

\begin{figure}[H]
\includegraphics[%
scale=0.8]{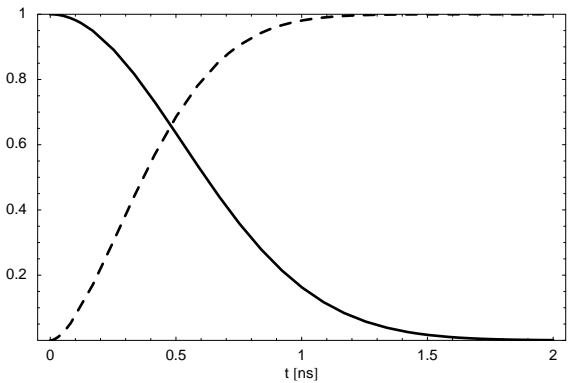}
\caption{Off-diagonal elements of the density operator in the spin
space showing the loss of coherence (solid curve) and the
entanglement between the spin and coordinates degrees of freedom
(dashed curve). Plotted with the typical values mentioned in the
text.}
\end{figure}
\begin{figure}[!ht]
\includegraphics[%
scale=0.8]{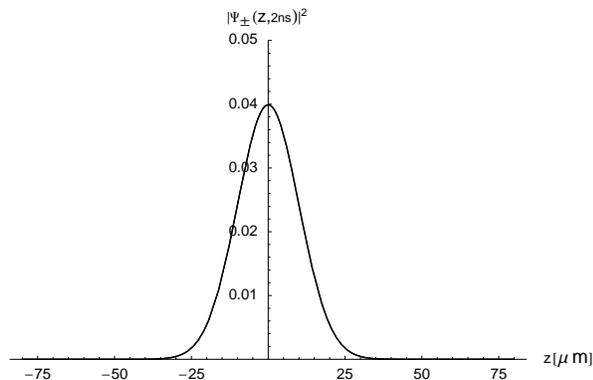}
\caption{The probability amplitudes for the spatial part of the
wave function at 2 ns, when  coherence is very small.}
\end{figure}
\begin{figure}[!ht]
\includegraphics[%
scale=0.8]{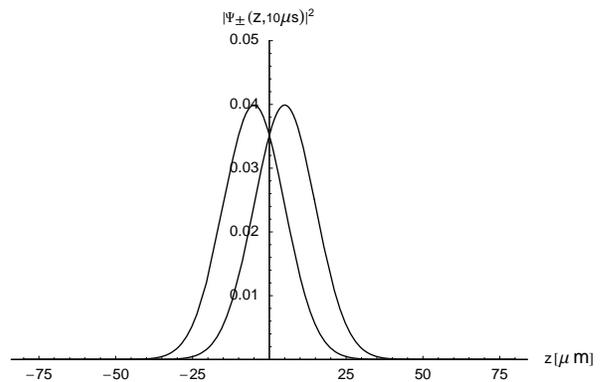}
\caption{Probability amplitudes for the spatial part of the wave
function at 10 $u$s, time interval beyond which  the packets
start to become separated.}
\end{figure}

Another interesting thing to look at is the behaviour of the
entanglement between the spin and  spatial coordinates. As we are
dealing with a global pure state we can use either the von Neumann
\cite{european j} or the linear entropy of one of the subsystems
as a measure of entanglement. We shall develop the latter in what follows.

Usually we have a bipartite system with the global state given by
\[|\psi\left(t\right)\rangle=\alpha\varphi_{+}\left(z,t\right)\left|+\right\rangle
+ \beta\varphi_{-}\left(z,t\right)\left|-\right\rangle \] \\
and as the coefficients $\alpha$ and $\beta$ vary the degree of
entanglement of the state of the system also changes. Two extremes
cases are the state of maximum entanglement, the Bell state, in
which the coefficients are both $1/\sqrt{2}$ and the separable
state (non-entangled) in which one of the coefficients is zero.

In our case we have a different physical situation since what is
varying is not the coefficients but the states themselves. In
other words, we are varying the states
$\varphi_{+}\left(z,t\right)$ and $\varphi_{-}\left(z,t\right)$
that start {}``parallel'' to one another and  become orthogonal as
time evolves. The entanglement between the spatial and spin
degrees of freedom can be given, in terms of the linear entropy of the
spin subsystem, by $E_{L}=1-\rho_{+-}^2(t)$. In this expression we
can see the close connection between entanglement and coherence;
as one of them increases the other is fated to decrease. This
behaviour can be observed in Fig. 4 where we have ploted $E_L(t)$
and $\rho_{+-}(t)$ in the typical SG experiment. As expected, the
entanglement vanishes at the beginning and increases as the states
become orthogonal, having as a consequence the loss of spin
coherence.

This example clearly shows the double role played by entanglement;
it allows for the creation of superpositions of macroscopic
objects but at the same time makes  subsystems lose their quantum
mechanical coherence.

For the sake of completeness, we should mention that one could
also think about tracing out the spin degrees of freedom and see
the coherence in the coordinate representation as an interference
pattern between the two beams in a double slit experiment. The
only "problem" here is that the spin degrees of freedom are always
orthogonal which makes the reduced density matrix in coordinate
space diagonal and destroy the possible coherence at any time.

Finally, since the present analysis makes  the connection between
entanglement and coherence clear only for pure states it would be
desirable to extend it to mixed states as well.

\section*{SUMARY}

We have given a simple example of the connection between
entanglement and coherence showing how the spin coherence is lost
in a SG experiment as the spatial and spin degrees of freedom
become entangled. We have also observed that, contrary to the
expectations, the spin coherence is lost much faster than the two
beams become clearly separated. It is worth commenting that we
have used the word decoherence with a somewhat different meaning
from that used in the current literature. Here we do not have a
real environment as the cause  of decoherence, and therefore are
referring to a possible case of reversible decoherence. Recovery
of coherence should be achieved simply by recombining the two
beams (See Ref.s \cite{ISG1,nosso,dugic} for more details on
this).

\begin{acknowledgments}

T. R. O., is grateful for FAPESP whereas A.O.C. acknowledges the support
from CNPq Conselho Nacional de Desenvolvimento Cient\'ifico e
Tecnol\'ogio and the Millennium Institute for Quantum Information.

\end{acknowledgments}

\end{document}